\documentstyle[aps,multicol,psfig]{revtex}

\newcommand{\be}{\begin{equation}}
\newcommand{\ee}{\end{equation}}
\newcommand{\ba}{\begin{eqnarray}}
\newcommand{\ea}{\end{eqnarray}}

\begin{document}

\title{
Entangling atoms in photonic crystals
}
\author{Martin Kon{\^o}pka$^{1}$ and 
Vladim{\'\i}r Bu\v{z}ek$^{1,2}$
}
\address{
$^{1}$ 
Faculty of Mathematics and Physics,
Comenius University, Mlynsk\'a dolina, 842 15 Bratislava, Slovakia\newline
$^{2}$
Institute of Physics, Slovak Academy of Sciences, D\'ubravsk\'a cesta 9,
842 28 Bratislava, Slovakia\newline
}
\date{18 January, 1999}
\maketitle
\begin{abstract}
We propose a method for 
entangling a system of two-level atoms in  photonic crystals.
The atoms are assumed to move in void regions of a photonic crystal.
The interaction between the atoms is mediated  either 
via a defect mode or via resonant dipole-dipole interaction. 
We show that these interactions can produce pure entangled atomic
states.
We analyze the problem with parameters typical for currently existing 
photonic crystals and Rydberg atoms. 
We show that the atoms can emerge from photonic crystals
in entangled states. 
Depending on the linear dimensions of the 
crystal and on their velocity of the entangled atoms can be separated
by tens of centimeters.
\end{abstract}


\vspace{-1truecm}

\begin{multicols}{2}

\section{Introduction}
\label{s_intro}

Quantum entanglement is one of the most remarkable feature of quantum
mechanics. Coherent control of the entanglement between
quantum systems attracts lot of attention mainly because of its potential
application in quantum information processing. Simultaneously, 
experimental investigation of the entanglement allows us to test
basic postulates of quantum mechanics and to answer fundamental
epistemological questions. These questions are related to the
original 
Gedanken experiment of Einstein, Podolsky and Rosen \cite{EPR} which
triggered 
discussions about non-locality of quantum mechanics and motivated 
experimental proposals 
to test whether quantum mechanics is the complete non-local theory.
The first experimental confirmation of the violation of Bell's 
inequalities \cite{Bell} has been done with the help of entangled photons
\cite{Aspect}.
A weak point of  experiments with photons is an 
insufficient control of  directions of   emitted photons and 
 small detectors efficiencies.
This problem should be removed in proposals where  
highly excited (Rydberg)
atoms are entangled.
Probably the first proposal of such an experiment is described in
Ref. \cite{Oliver}. Other proposal have been presented in Refs.
\cite{atoms}. 
Authors of these schemes proposed 
techniques how to create 
entangled atoms in microwave single-mode cavities.
Recently, controlled entanglement between atoms separated apr. by $10$ mm
  interacting
with an electromagnetic field in  a high-$Q$ cavity 
has been experimentally realized \cite{Hagley}.
In addition, trapped ions have been created in entangled states
\cite{ion}.

In this paper we propose   a simple scheme for entangling atoms  
in photonic crystals. We remind us that 
photonic crystals are artificially created three-dimensional periodic
dielectric materials which exhibit a frequency gap or several gaps 
in spectrum of 
propagating electromagnetic (EM) waves \cite{Yablonovitch1987,John1987}. 
An EM wave with its frequency from the
gap can not propagate in the structure in any direction.
Photonic crystals operating at microwave frequencies were
successfully created in laboratories \cite{Yablonovitch1991a}. They consist
of a solid dielectric and empty regions. 
The periodicity of a photonic crystal can be destroyed by removing or adding a
piece of material which creates a defect EM mode in the structure.
This mode is spatially localized around the region of the defect. 
The frequency of the mode and the spatial modulation of its electric field
amplitude depends on  properties of the defect
\cite{Yablonovitch1991b,McCall,Winn,Smith}. 
It means that one can
adjust  parameters of the defect mode by creating a suitable defect in
the crystal. In particular, the spatial dependence
of the mode amplitude can be adjusted to particular needs.
In quantum optics, defect modes in photonic crystals can be used similarly 
as high-$Q$ single-mode cavities \cite{Yablonovitch1994,Meschede}. 
The quality factor of a single mode in a metallic cavity can be of 
order of $10^8$ or more and similar values can be reached for a single
defect mode in a photonic crystal \cite{Yablonovitch1994}.
Today three-dimensional photonic crystals are available only at
microwave frequencies. They can be used for experiments with Rydberg atoms, 
similarly as microwave cavities.

In this paper we consider two interactions via 
which one can produce entangled atoms.
Firstly, we show that at
it is possible to generate entangled atoms
without a defect mode, using the action 
of the resonant dipole-dipole interaction (RDDI) \cite{John1991,John1995} 
mediated by off-resonant modes of photonic band continua.
Secondly, we explore the scheme in which 
the atoms become mutually entangled due to the interaction with the 
defect-field mode.

The paper is organized as follows: 
Basic features  of the proposed  setup 
are described in Section \ref{s_setup}.
In Section \ref{s_RDDI} we discuss how the atoms in photonic crystals
can be entangled via the resonant dipole-dipole interaction.
In Section \ref{s_EPR} we study in detail the entanglement of atoms
which interact with a single defect mode in the photonic crystal.
In Section \ref{s_last} we conclude the paper with some remarks.


\section{Setup of the scheme}
\label{s_setup}

We consider two mechanisms via which  a system of identical 
atoms can be entangled in photonic crystals. 
We assume that the  atoms are modeled by two-level systems having their 
transition frequencies in a photonic bandgap (PBG).

The first mechanism is the resonant dipole-dipole interaction (RDDI) 
mediated by  off-resonant modes of the
photonic-band continua [see the Hamiltonian (\ref{Hameff})]. 
This interaction has been analyzed
in detail  by Kurizki
\cite{Kurizki1990} and John and Wang \cite{John1991} as well as 
by John and Tran Quang \cite{John1995}. 
These authors have considered a system of two-level atoms. They 
have shown that if one of the atoms is excited and the other one is in its
ground state, then they can exchange excitation in spite of the fact that 
their transitions frequencies are in
a PBG and spontaneous emission is nearly totally suppressed.
The RDDI can  be understood as an energy exchange via
localized field \cite{John1991}. 
This light tunneling (or photon-hopping
conduction) can be very efficient when 
the distance between  the atoms is much smaller than  the light wavelength. 
The RDDI can  occur either in a free space or in a cavity.
However, in a free space the excitation is irreversibly radiated into the 
continuum of the field modes after a very short time (given by Fermi's
Golden rule) and the entanglement between the atoms is deteriorated
rapidly.

The second mechanism is due to 
an excitation exchange via a defect mode which is
resonant (or nearly resonant) with the atoms. This type of interaction
explicitely involves a quantized defect mode and is described by
the Hamiltonian (\ref{HamDef}).

These two interactions can also
occur simultaneously. As we will see, the second mechanism is much 
more efficient and allows a coherent control over the process of entanglement. 
The first mechanism can be neglected in many
cases, especially when the atoms have their transition frequencies 
near the center of a wide PBG 
and their distance is not much smaller than the 
wavelength of the resonant light.

\begin{figure}[t]
\centerline{\psfig{height=7.5cm,file=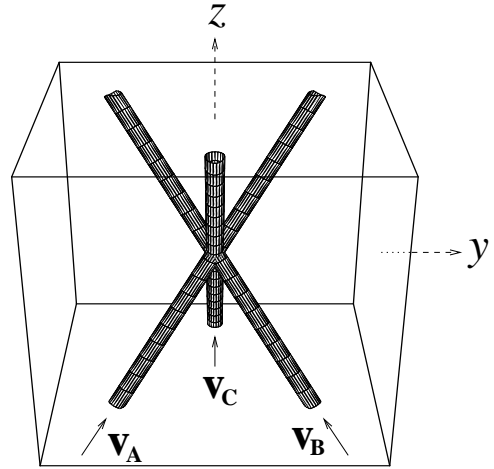}}
\begin{narrowtext}
\caption{
A schematic description of the physical situation. We
note that a particular geometry of the scheme and the type of the photonic
crystal are not essential. The important feature
of the crystal geometry is that it has straight ``tunnels'' so that atoms can
traverse across the crystal.
For concreteness, we choose the crystal with the geometry proposed 
by Yablonovitch {\it et al.} (references in Section \ref{s_setup}).
We display only three of many cylindrical holes in the crystal.
The cylindrical holes are drilled 
at the angle $\Theta = 35.26^\circ$ with the vertical axis. 
We assume (except Section \ref{s_RDDI}) that there is a defect of
the crystal periodicity near the region where the holes are crossing.
This defect is responsible for 
 a single defect mode localized in the center of the
crystal. The frequency of the defect mode lies inside a wide
photonic bandgap of the crystal.
Linear dimensions of the defect-mode region are comparable with the lattice
constant of the structure. 
The atoms $A$, $B$ (and $C$, if
needed) are injected into the three holes at the bottom side of the crystal 
at (approximately) the same time
and with suitably adjusted velocities. 
The atoms are assumed to be two-level systems
with their transition frequencies equal to the defect-mode frequency. We
assume that transitions including other atomic levels can be neglected.
The atom $A$ is injected in its upper level $|e_A\rangle$ 
and the atom $B$ and $C$, 
in their lower levels $|g_B\rangle$ and $|g_C\rangle$. The initial state
of the defect mode is vacuum state.
The states of the atoms are detected at the exit from the crystal.
We consider the following numerical values for the setup. The crystal is a
cube of the side $L \approx 20$ cm. 
The frequencies of
the defect mode and the atomic transition are 
$\omega_0/(2\pi) = \omega/(2\pi) = 21.50651$ GHz, i.e. the same as
transitions used in experiments with microwave cavities.
This frequency lies inside the wide photonic bandgap if the 
crystal is made from dielectric with refractive index (at microwaves) $3.6$,
volume filling fraction is $78 \%$ and the side of an elementary cube is 
$a \approx 16.3$ mm. We use this value of $a$ to calculate the 
parameter $k = \pi/a$ [see Eq.(\ref{shape})].}
\label{fig1}
\end{narrowtext}
\end{figure}

In what follows 
we describe the basic setup of the proposed experiment in the case 
when the atoms interact only via the defect mode. 
Let us assume that one of the three atoms (let say the atom $A$)
is  prepared initially 
in its excited state while the other two atoms ($B$ and $C$) are initially
in their  ground states (see Fig.\ref{fig1}). 
After the preparation the atoms are  injected into
cylindrical void regions of the crystal. We consider the photonic crystal of
the geometry designed by Yablonovitch {\it et al.}
\cite{Yablonovitch1991a,Yablonovitch1994}
although  other appropriate geometries can be used as well. 
The void cylinders intersect at the center of the crystal.
The defect-field mode (located near the center of the crystal) is initially 
prepared in its vacuum state.
Firstly the atoms propagate freely in the void cylinders 
outside the defect-field (this is due to the fact that the
 transition frequencies of the atoms lie inside the wide PBG).  
When the atoms enter the defect region 
they start to interact with the single defect-field mode. And then again,
after they leave   the defect region they   evolve freely. 
If the exited (ground) state
of atom $j$ ($j=A,B,C$) is denoted as $|e_j\rangle$ ($|g_j\rangle$) 
and the $n$-photon
state of the single mode 
defect field is denoted as $|n\rangle$ then 
the initial state of the system under consideration can written as
\be
|\Psi(0)\rangle = |e_A\rangle\otimes |g_B\rangle\otimes |g_C\rangle\otimes
|0\rangle \equiv |e_A, g_B, g_C, 0\rangle.
\label{initstate}
\ee
When we assume that in the defect region the atom-field interaction
is governed by the Hamiltonian in the dipole and rotating 
wave approximations (see below)           
then the final state of the system reads 

\end{multicols}
\widetext
\be
|\Psi(t)\rangle = a(t) |e_A, g_B, g_C, 0\rangle + 
b(t) |g_A, e_B, g_C, 0\rangle + 
c(t) |g_A, g_B, e_C, 0\rangle + 
\gamma(t) |g_A, g_B, g_C, 1\rangle,
\label{finalstate2}
\ee
\begin{multicols}{2}
\noindent
where $t$ is the time at which we detect the internal states of the atoms
at the exit of the crystal.  
The final values of the amplitudes $a$, $b$, $c$ and $\gamma$
depend on a particular setup of the experiment including the coupling 
parameters and velocities of the atoms.
For completeness of the description we specify trajectories ${\bf r}_j(t)$
of the three 
atoms which can move along the axes of the three void regions 
\be
{\bf r}_j(t) = {\bf r}_j(0) + {\bf v}_j t;\qquad j=A,B,C
\label{trajectoryj}
\ee
with the vectors ${\bf r}_j(0)$  and ${\bf v}_j$ specified by their
components as
\ba
{\bf r}_A(0) &=& \frac{L}{4} \left\{\tan\Theta, -\sqrt{3} 
\tan\Theta, -2 \right\},
\nonumber
\\
{\bf v}_A & =& \frac{v_A}{2} \left\{- \sin\Theta, \sqrt{3} 
\sin\Theta,2\cos\Theta\right\},
\label{trajectoryA}
\ea
for the atom $A$. While for the other two atoms ($B$ and $C$) we have
\ba
{\bf r}_B(0) &=& 
\frac{L}{4} \left\{
\tan\Theta, \sqrt{3} \tan\Theta, -2 \right\};
\nonumber
\\
{\bf v}_B &=& \frac{v_B}{2} \left\{- \sin\Theta, 
-\sqrt{3} \sin\Theta, 2\cos\Theta\right\};
\label{trajectoryB}
\ea
and
\ba
{\bf r}_C(0) &=&
\frac{L}{2} \left\{-
\tan\Theta, 0, -1\right\};
\nonumber
\\
{\bf v}_C &=& v_C \left\{\sin\Theta, 0, \cos\Theta\right\}.
\label{trajectoryC}
\ea
Here we assume  the origin of the coordinates  in the center
of the cube crystal with the side of the length $L$; $\Theta$ is the 
angle between the axes of the cylinders and the $z$ direction.


\section{Entanglement via resonant dipole-dipole interaction}
\label{s_RDDI}

In this Section we consider just two identical  atoms 
($A$ and $B$) which move in the crystal as it is described above.
Here we assume that there is no defect mode in the crystal.
The atoms move inside the crystal with 
constant velocities. The recoil effect due to interaction with
electromagnetic field is  neglected because the 
atoms are relatively heavy particles. 
The interaction between the atoms and the electromagnetic field modes
inside the crystal is described by the Hamiltonian in the
electric-dipole approximation
\ba
H &=& \hbar \omega \sum_{j=A,B} \sigma_z^j
+\hbar \sum_\lambda  \omega_\lambda a^\dagger_\lambda a_\lambda 
\nonumber
\\
&-&
\frac{1}{\epsilon_0}{\bf \mu}(A) \cdot {\bf D}({\bf r}_A) - 
\frac{1}{\epsilon_0}{\bf \mu}(B) \cdot {\bf D}({\bf r}_B),
\label{Ham}
\ea
where 
$a_\lambda$ and $a^\dagger_\lambda$ are the annihilation and creation
operators of the field mode labeled by $\lambda$, ${\bf D}({\bf r})$ is the
transverse displacement-field operator and ${\bf \mu}(A)$ and ${\bf \mu}(B)$ 
are the atomic dipole operators.
When the
atomic transition frequencies are far from abrupt changes in the density 
of modes the Hamiltonian (\ref{Ham}) can be approximated as 
(for more details see \cite{John1995}) 
\be
H_{\rm eff} = \hbar \omega \sum_{j=A,B} \sigma_z^j + 
\hbar \left( J_{AB} \sigma_+^A \sigma_-^B 
+ J_{BA} \sigma_-^A \sigma_+^B\right),
\label{Hameff}
\ee
where 
$\sigma_\pm^x$ are raising and lowering operators of the atoms ($x=A,B$)
and 
$J_{AB}$ is a matrix element for
the effective description of the RDDI \cite{John1991}. 
For qualitative estimations, we will use $J_{AB}$ evaluated under
the  assumption
that density of electromagnetic modes is that of a free space.
In this case we find 
(for more details see \cite{Craig})
\end{multicols}
\vspace{-0.5cm}
\noindent\rule{0.5\textwidth}{0.4pt}\rule{0.4pt}{\baselineskip}
\widetext
\be
\hbar J_{AB} = \mu^{ge}_i(A) \mu^{eg}_j(B) \frac{1}{4 \pi \epsilon_0 R^3}
[(\delta_{ij} - 3 \hat{R}_i \hat{R}_j)(\cos k_A R + k_A R \sin k_A R) 
- (\delta_{ij} - \hat{R}_i \hat{R}_j) k_A^2 R^2 \cos(k_A R)],
\label{JAB}
\ee
\begin{multicols}{2}
\noindent
where $R$ is the distance between the atoms, $k_A \equiv \omega/c$, 
 $\mu^{eg}$ is the absolute value of the atomic
dipole matrix element, and $\hat{R}_i$ are the components of the unit
vector starting at the position of the atom $A$ and oriented towards
the atom $B$. We assume summation over repeated indeces.
We stress that the above expression for $J_{AB}$
is valid in a free space, but  in the limit 
$R \ll  \lambda$ it can also be applied for photonic crystals
\cite{John1991},
i.e. it can be also used for an order-of-magnitude 
description of the  RDDI effects in photonic crystals. 
These effects are
most important in the regime $R \ll \lambda$ when the free-space expression 
is valid also in photonic crystal.
We will apply the Hamiltonian (\ref{Hameff}) with $J_{AB}$ given
by Eq.(\ref{JAB}) also for
description of propagation of the atoms in the crystal in the case when   
$R\ge L$. Even though the expression for $J_{AB}$ given by Eq.(\ref{JAB})
is not precise
it provide us with a rather good picture
of the RDDI effect.
We note that 
in order to 
find more appropriate expression for  $J_{AB}$ 
we would have to know the electromagnetic
eigenmodes  for the three-dimensionally periodic
structure and the corresponding derivation of $J_{AB}$ is very complicated.

\begin{figure}[t]
\centerline{\psfig{height=7.5cm,file=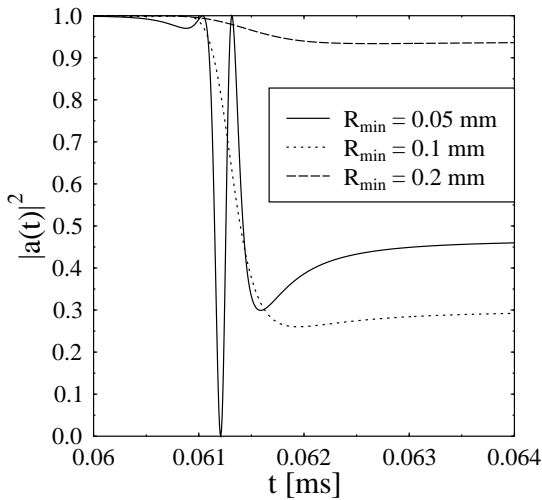}}
\begin{narrowtext}
\caption{
The time evolution of the population of the upper level of the 
atom $A$ if the atoms interact according to effective Hamiltonian
(\ref{Hameff}). 
The atoms move along the trajectories specified in Section \ref{s_setup}.
Three curves correspond to three different values of $x_A(0)$ specified by 
the minimal distance $R_{\rm min}$ of the atoms during the passage.
Both atomic dipoles are oriented in the $x$-direction.
We evaluate the interaction only during 
the time interval when the atoms move near the center of the crystal 
in the cubic region of the side $2$ cm.
The atomic velocities are $v_A = v_B = 200$ m s$^{-1}$ and 
$\mu^{eg}/e \approx 6.72\ 10^{-7}$ m, where $e$ is the proton charge.
}
\label{fig2}
\end{narrowtext}
\end{figure}

In what follows we study the time evolution of the atoms
initially prepared in the state 
$|\Psi(0)\rangle = |e_A, g_B\rangle$ which is governed
by the effective 
Hamiltonian (\ref{Hameff}) with time-dependent $J_{AB}$
(which is due to the fact that the atoms are moving through the crystal).
We show that the RDDI can in principle be used
for controlling the entanglement between atoms.
We have solved the corresponding Schr{\" o}dinger equation numerically. 
We have used  parameters typical
for Rydberg atoms and currently existing photonic crystals. 
In Fig.\ref{fig2} we plot results for the time-dependent atomic populations.
We have chosen the atomic trajectories similarly as it is specified in the 
previous section but we added a small value ($0.05 - 0.3$ mm) to the initial
$x_A(0)$ coordinate so that the trajectory of atom A is parallel but not 
identical with the axis of the cylinder. This prevents the collision of the
atoms. The velocities of both atoms are  $200$ m s$^{-1}$. 

Taking into account that the physical conditions are chosen such that the
electromagnetic field is adiabatically eliminated from the interaction
[see the effective Hamiltonian (\ref{Hameff})] the two atoms due to the
unitarity of the evolution remain in a pure state
$ 
|\Psi(t)\rangle_{AB} = a(t) |e_A, g_B\rangle + 
b(t) |g_A, e_B\rangle 
$
with the amplitudes $a(t)$ and $b(t)$ which depend on the RDDI.
From here it follows that due to the RDDI the two atoms become
entangled. The degree of entanglement in the present
case can be quantified with the help of the von Neumann entropy
$S=-{\rm Tr}[\hat{\rho}\ln \hat{\rho}]$ of each individual atom
for which we have $S=-|a(t)|^2\ln |a(t)|^2 -|b(t)|^2\ln |b(t)|^2$
where $|a(t)|^2=1-|b(t)|^2$. In other words the degree of the entanglement
depends on the population of internal levels 
of the atoms and highest degree of entanglement is attained for
$|a(t)|^2=|b(t)|^2=1/2$.

As seen from Fig.~\ref{fig2}
the population of the excited state of the atom $A$
depends on
the minimal distance $R_{\rm min}$ between the atoms during the passage 
through the crystal.
From our numerical investigation it follows
that the atoms are most entangled  
for $R_{\rm min} \simeq 0.05$ mm.
However we note that with  present techniques the controle over the
position of atoms in the configuration considered here is 
about $\pm 1$ mm \cite{Hagley}.
Consequently, the RDDI is not very suitable for a coherent controle
of entanglement between atoms in photonic crystals.
In the following Section we consider entanglement via a defect mode when the
currently available precision control is sufficient.


\section{Entanglement via a defect mode}
\label{s_EPR}
Let us consider the interaction of the atoms with a 
single defect-field mode in  the dipole and 
the rotating-wave approximations. 
We assume that the distance between 
the atoms is always sufficiently large so that they do not
interact via RDDI.
The corresponding Hamiltonian can be written as
\ba
H &=& \hbar \omega \sum_{j=A,B,C} \sigma_z^j  + \hbar \omega_0 a^{\dagger} a 
\nonumber
\\
&+&\hbar\sum_{j=A,B,C} \left[G({\bf r}_j) \sigma_+^j a + G^*({\bf r}_j)
\sigma_-^j a\right],   
\label{HamDef}
\ea
where $\omega_0$ is the mode frequency (which we assume to be equal to
the atomic transition frequency  $\omega$), 
$\sigma_{\pm}^j$  are atomic raising and 
lowering operators and ${\bf r}_A$ and ${\bf r}_B$ are the positions of the
atoms.
The position dependence of the coupling parameters $G({\bf r}_j)$
can be expressed as  
\be
G({\bf r}_j) = G_0\  {\bf \epsilon} \cdot {\bf {\cal D}}_j\ f({\bf r}_j),
\label{coupling}
\ee
where $f({\bf r})$ is the field-mode amplitude at the position ${\bf r}$, 
${\bf \epsilon}$ is the electric-field 
polarization direction of the defect mode and ${\bf {\cal D}}_j$ is a unit
vector in the direction of the atomic dipole matrix element of the atom $j$.
It is known that the spatial dependence of a defect-mode amplitude is a
function which oscillates and decays exponentially \cite{McCall}. 
A particular profile of the spatial dependence 
of the defect mode can be  adjusted via a properly generated 
defect of the periodicity. 
A rigorous calculation
of the electromagnetic field in the presence of a defect in a $3$D photonic
crystal can be a difficult task. 
In this paper we use
a model profile of the spatial dependence of the electric field. Similar
profiles have already been created in existing photonic crystals
\cite{Yablonovitch1991b,McCall,Winn,Smith}.
We note that for the purpose of the proposed experiment
a complete information about the mode shape is not needed. The results
of the experiment depend only on the shape along the trajectories of the
atoms. In what follows we use the profile 
\be
f({\bf r}) = \exp{\left[-\frac{|{\bf r} - {\bf R}_0|}{R_{\rm def}}\right]}
\sin ({\bf k} \cdot {\bf r} + \Phi),
\label{shape}
\ee
where
${\bf R_0}$ is the position around which the mode is localized, $R_{\rm def}$ 
is a parameter (defect-mode radius) describing the rate of the 
exponential decay of the 
mode envelope, $\Phi$ is a phase factor and ${\bf k}$
is the parameter describing spatial oscillations of the field mode.
We chose its 
magnitude to be $k = \pi/a$ where
$a$ is the value of the side of an elementary cubic cell in the photonic
crystal.
We consider values of the constant $R_{\rm def}$ comparable
with $a$.
We  estimate the value of the coupling constant 
$G_0$ from  microcavity experiments \cite{Meschede} 
\be
G_0 = \sqrt{\frac{V_{\rm cav}}{V_{\rm eff}}} \Omega ,
\label{G0}
\ee
where $V_{\rm cav}$ is the modal volume of the microcavity mode, $V_{\rm
eff}$ is the effective modal volume of the defect mode and $\Omega$ 
is the vacuum Rabi frequency in the microwave experiment.
The numerical values are \cite{Meschede}: $V_{\rm cav} = 11.5\ {\rm cm}^3$ 
and $\Omega = 43$\ kHz. When we consider the transitions between 
levels $63 P_{3/2}$ and $61 D_{3/2}$ of
Rubidium atoms, then  the atomic transition frequency is  
$\omega/(2\pi) = 21506.51$\ MHz.
Finally, the effective modal volume can be approximated as
\be
V_{\rm eff} = \frac{4}{3} \pi (2 R_{\rm def})^3.
\label{Veff}
\ee
Because the atoms are moving the coupling parameters depend on time [in what
follows we will use the notation $G_j(t)$]
We consider positions of the atoms given by Eqs.(\ref{trajectoryA}) and 
(\ref{trajectoryB}). In some cases we add a small value to $x_A(0)$ 
given by (\ref{trajectoryA}) to
prevent the atoms to collide in the center of the crystal.
Details of the geometry of the proposed experiment are given in Section
\ref{s_setup} and in Fig~.\ref{fig1}.

Once we have specified all model parameters we can solve the
Schr{\" o}dinger equation for the system which is supposed to be 
initially prepared in the state
$|\Psi(0)\rangle = |e_A, g_B,g_C, 0\rangle$. 
Due to the fact that the number of excitations is an integral of motion
in the present case the state vector at time $t>0$ has the
form (\ref{finalstate2}) and the corresponding Schr{\" o}dinger
equation can be rewritten into a set of a system of linear
differential equations. 
These equations can be solved analytically
for time-independent coupling constants $G_j(t)$ which is not our case.
Therefore
we have to integrate the equations numerically.

\subsection{One atom}
We start our discussion  with a problem when just a single atom
(let say the atom $A$)  
passing through the crystal is considered. We assume that the
atom is on resonance with the defect mode
(i.e., $\omega = \omega_0$).
\end{multicols}
\vspace{-0.5cm}
\noindent\rule{0.5\textwidth}{0.4pt}\rule{0.4pt}{\baselineskip}
\widetext
\begin{figure}[t]
\centerline{\psfig{height=7.5cm,file=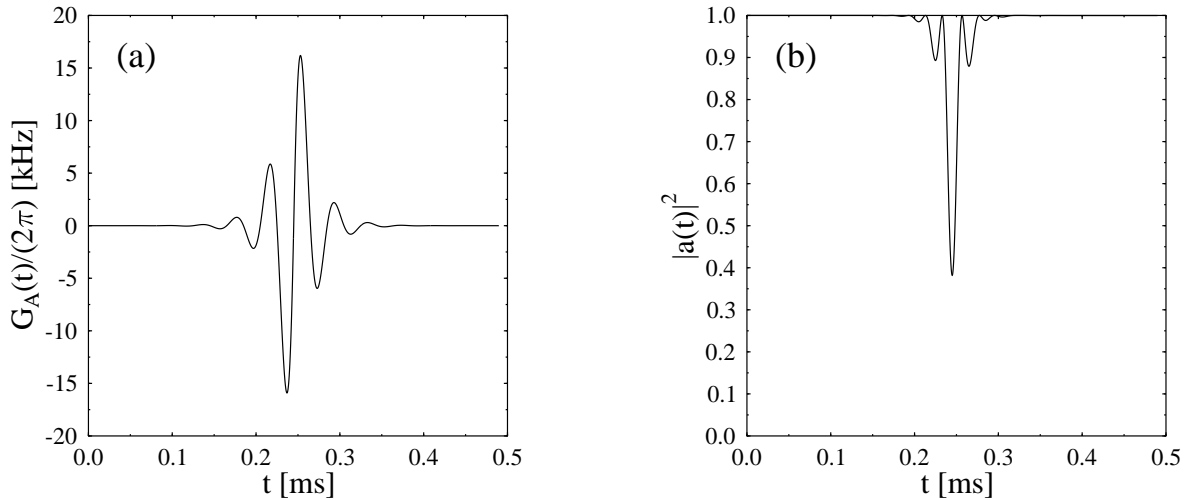}}
\caption{
(a) The time 
dependence of the coupling   $G_A(t)$ 
between the defect mode and  the
single atom $A$ when it moves along the axis of the cylinder with the velocity 
$v_A = 500\ {\rm m s}^{-1}$. 
The mode position and the geometry is given by the parameters 
${\bf R_0} = {\bf 0}$, $\Phi = 0$ rad and $R_{\rm def} = 10$ mm. The
parameter ${\bf k} = (0,0,\pi/a)$ where $a \approx 16.3$ mm. The atomic
dipole is oriented in the $x$ direction (same as the field polarization).
The integral (\ref{int_G}) in this case is equal to zero.
Consequently, the atom at the exit of the crystal is again in its
initial state. We plot the time evolution of the population of the
exited level of the atom in Fig.b.
}
\label{fig3}
\end{figure}
\begin{multicols}{2}
\noindent
This corresponds 
to the Jaynes-Cummings model \cite{JCM}  with
a time-dependent coupling constant. The general solution of this model 
for real coupling
parameter was found Sherman et al.\cite{Sherman}. With the initial condition
$|\Psi(0)\rangle_ = |e_A, 0\rangle$ the solution can be expressed as
\ba
|\Psi(t)\rangle &=& \cos\left[\int_0^t G_A(t') dt'\right] |e_A, 0\rangle
\nonumber
\\
&-& i \sin\left[\int_0^t G_A(t') dt'\right] |g_A, 1\rangle.
\ea
This implies for the atomic excitation
\be
P_{\rm e}^{(A)}(t) = \cos^2\left[\int_0^t G_A(t') dt'\right].
\ee
In the case of the defect mode with linear dimensions 
much smaller
than the side of the crystal we can use the approximation
\be
\int_0^t G_A(t') dt' \approx \int_{-\infty}^{\infty} G_A(t') dt'.
\label{int_G}
\ee
We note that this integral for a given choice of the profile function 
[see Eq.(\ref{shape})]  with
the phase  of the field mode $\Phi = 0$
equals to zero. This means that the atom exits the crystal in
the same state as it entered it. Obviously the defect mode also
remains in its initial (vacuum) state.
In Fig.\ref{fig3}a we plot the time dependence of 
the coupling constant between the 
atom $A$ and the defect mode. While 
in Fig.\ref{fig3}b we present the  
time dependence of the corresponding excited-state probability. It is 
assumed that the
defect is located at the center ${\bf R}_0 = {\bf 0}$ of the crystal.
The other parameters are chosen such that 
$\Phi = 0$ rad , ${\bf k} = (0,0,k)$, ${\bf {\cal D}}_A = {\bf \epsilon}
= (1,0,0)$ [see Eqs. (\ref{coupling}),(\ref{shape})]. We assume that 
 the atom moves along the
axis of the cylindrical cavity. The velocity of the atom is chosen to be
$v_A = 500 {\rm m s}^{-1}$. 
From Fig.~\ref{fig3}a we clearly see that the atom on its way through
the crystal interacts with the defect mode just around the center of
the crystal. 
The other important feature is seen from Fig.\ref{fig3}b, i.e. 
The atom is transiently entangled with the defect mode in the center 
of the crystal.
Nevertheless it leaves the crystal in a pure (unentangled) state.
This effect of ``spontaneous'' disentanglement of the atom from the
defect mode is very important when we consider creation of pure entangled
state of two atoms.

\subsection{Two atoms}

\begin{figure}[t]
\centerline{\psfig{height=13.0cm,file=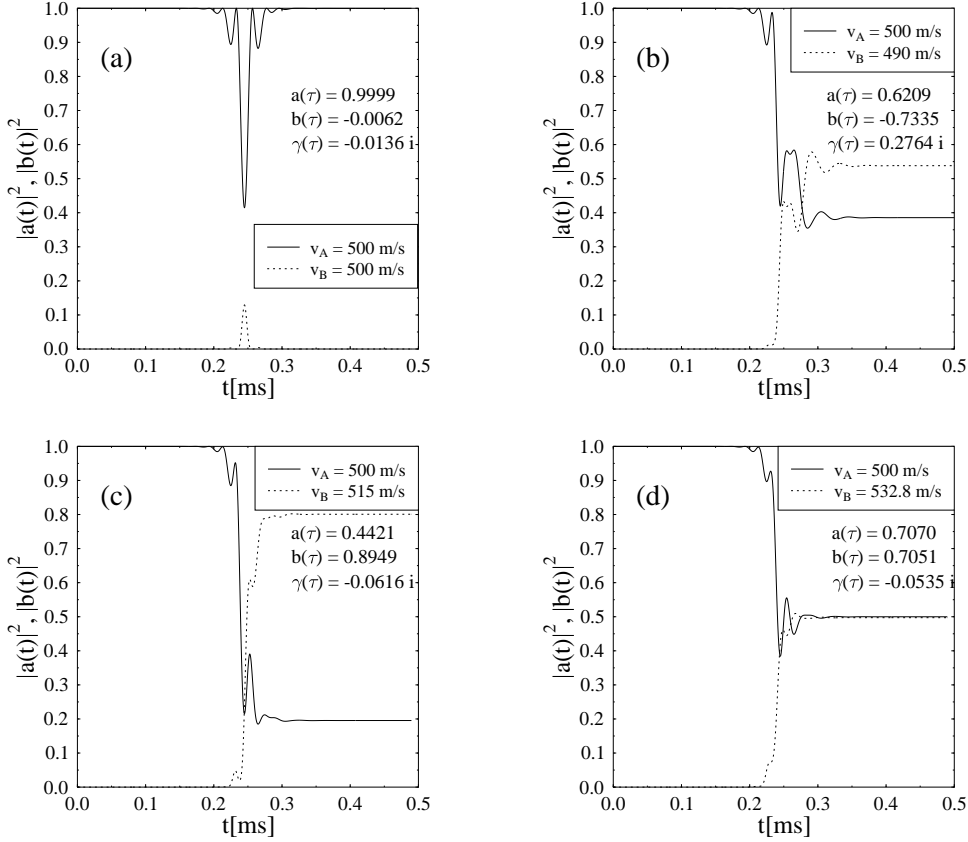}}
\caption{
The time evolution of the populations of excited levels of the
atoms $A$ (solid line) and $B$ (dashed line). 
The atom $A$ enters the crystal in the
excite state, while the atom $B$ is initially in the ground state.
The atoms are injected into
the crystal at the same time (see the setup presented by 
Fig.\ref{fig1}) with the velocity of the atom $A$ being
$v_A = 500\ {\rm m s}^{-1}$ and $R_{def}=10$mm.
The four plots corresponds to four
different velocities of the atom $B$ (their numerical values
are shown in figures).
In the case when the two atoms are assumed to have equal velocities
(Fig~a)  we add a small value ($0.3$ mm) 
to the initial position $x_A(0)$
[see the expression  (\ref{trajectoryA})] 
to avoid the collision of the atoms in the center of the crystal.
The other parameters are chosen same as in Fig.\ref{fig3}.
We write final values of the probability amplitudes into figures. 
From Fig.~(d) we see that for properly chosen velocities of the
atoms and times large enough (i.e. the atoms have already left
the crystal) 
the defect mode is approximately in the vacuum
state (it is disentangled from the atoms) and the atoms are
prepared in a pure superposition state $|\Psi\rangle\simeq(|e_A, g_B\rangle+
|g_A, e_B\rangle)/\sqrt{2}$.
}
\label{fig4}
\end{figure}

Let us consider a situation when two atoms interact 
with the same defect mode as in the previous case.
The atoms have their dipoles oriented along the
direction $\epsilon$ of the electric-field polarization.
The velocity of the atom $A$ is $500 {\rm m s}^{-1}$.
The time evolution of the  corresponding atomic
populations  for various velocities
of the atom $B$  are plotted in Fig.~\ref{fig4}.

Firstly we consider both atoms to have the same velocity 
(see Fig.~\ref{fig4}a). In this case 
we assume that the atom $A$ is displaced from the axis of the cylindrical hole 
through which it flies [i.e. we add   $0.3$ mm to $x_A(0)$
given by (\ref{trajectoryA})] to avoid the influence of the RDDI between the
atoms and their collision.
We see that the atoms strongly interact with the field in the region of
the defect. However, after the interaction the initial state of the
system is approximately restored (see the ``stationary'' values of the
probability amplitudes $a(\tau)$, $b(\tau)$ and $\gamma(\tau)$ which 
are displayed in the figures).
It is interesting to compare Fig.~\ref{fig3}b with
Fig.~\ref{fig4}a to see how the time evolution of the population
of the atom $A$ is  modified by the presence of the additional atom $B$.
We see that for the given set of parameters the presence of the atom
$B$ does not influence the dynamics of the atom $A$ significantly.

Now we will study how the level population depends on the 
velocity of the atom $B$. 
From Fig.~\ref{fig4} we see that for properly chosen
velocity the interaction between the atoms mediated by the defect
field can be pronounced. For instance, from Fig.~\ref{fig4}b
(here $v_B=490 {\rm m s}^{-1}$)
we see that not only the excitation of the atom $B$
can be higher than the population of the atom $A$, but also
the defect mode becomes partially excited and entangled with
the atomic system.

When the atom $B$ has the velocity $v_B=515 {\rm m s}^{-1}$ 
(see Fig.~\ref{fig4}c) then the defect mode in the stationary limit
is in the vacuum state [$\gamma(\tau)\simeq -0.0616 i$] 
and is (with high precision) completely disentangled from the
atomic system. It is interesting to note that in this particular situation
the defect mode mediates transfer of most of the excitation from 
the atom $A$ to the atom $B$.

Let us assume now the velocity of the atom $B$ to be 
$v_B=532.8 {\rm m s}^{-1}$ 
(see Fig.~\ref{fig4}d).
In this case the defect mode in the stationary limit
is again 
in the vacuum state and is completely disentangled from the
atomic system. Moreover the amplitudes $a(\tau)$ and $b(\tau)$ are
in this case almost equal, which means that the atoms at the exit from
the crystal are in the state $|\Psi\rangle = (|e_A, g_B \rangle
+|g_A, e_B \rangle)/\sqrt{2}$, i.e. they are prepared in a pure maximally
entangled state.

In the cases presented in Fig.~\ref{fig4} 
the phase factor $\Phi$ of the
defect mode is set to zero so that the integrals of the coupling constants
$G_A(t)$ and $G_B(t)$
over the trajectories of the atoms are equal to 
 zero. The defect-mode radius
$R_{\rm def} = 10$ mm. We have also studied the dynamics for other
values of $\Phi$, when 
the integrals of the coupling constants differ from zeros. 
In this case the disentanglement of the defect mode and the atoms
is not so well pronounced, i.e. the defect mode becomes excited.
We have also found a general feature: 
If the integrals of coupling
constants are zeros and the coupling constants are small enough then
the defect mode after the interaction is left in the vacuum state.
However, if we increase the couplings (by decreasing the mode volume $V_{\rm
eff}$) 
the defect mode can be left in an excited state [i.e. $\gamma(\tau)\neq 0$;
see the expression for the state vector (\ref{finalstate2})].
Consequently, the atoms are left in a mixture state.

We have also analyzed the situation when the defect mode is not located
directly at the center of the crystal. Moreover we have assumed that
$\Phi\neq 0$. It can be shown that even in this
case it is
 possible to find a value $v_B$ at which the atoms 
exit the crystal in a nearly pure maximally entangled state.

\end{multicols}
\vspace{-0.5cm}
\widetext
\begin{multicols}{2}

\subsection{Three  atoms}
\label{s_three}

\begin{figure}[t]
\centerline{\psfig{height=13.0cm,file=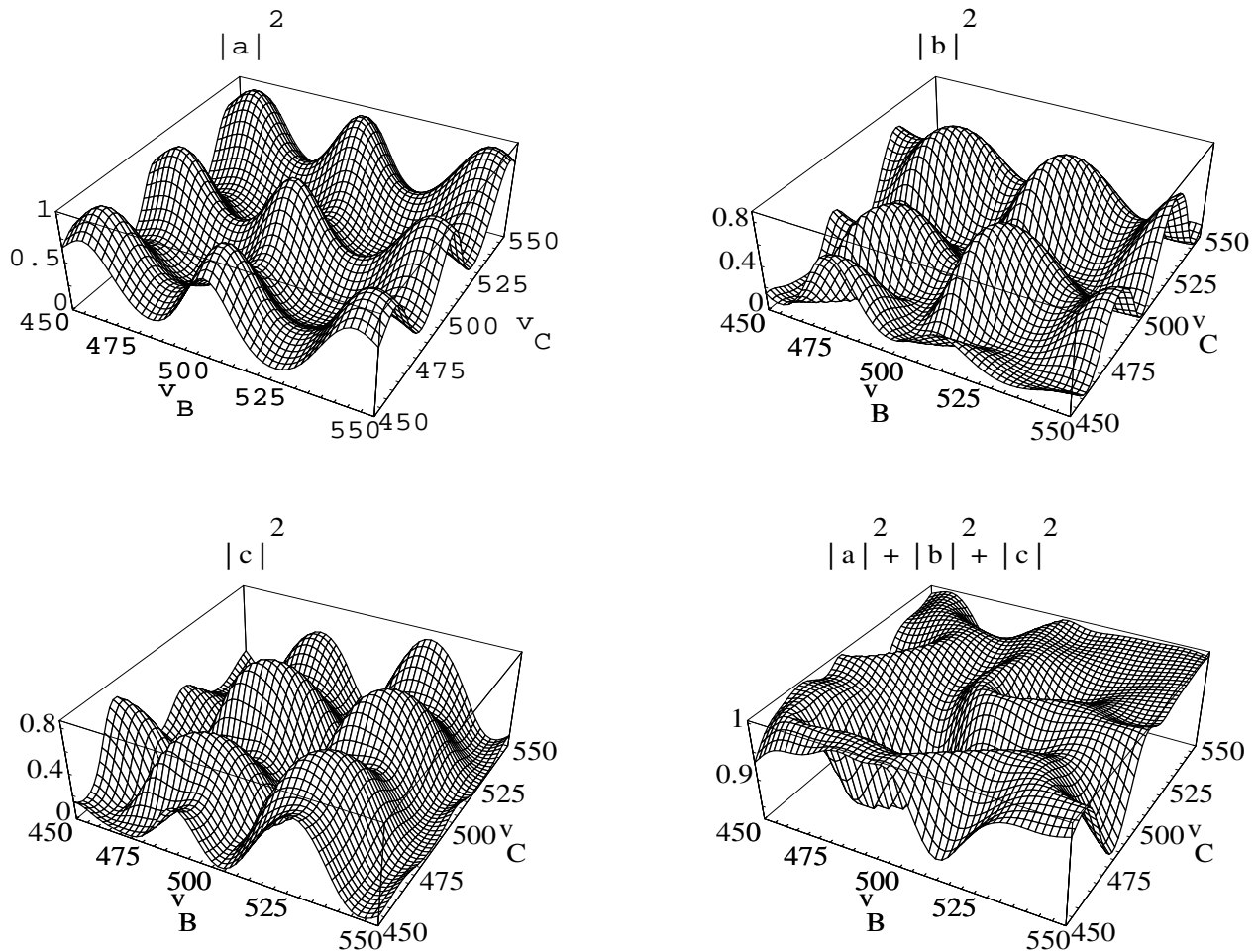}}
\caption{
Final atomic populations of the three atoms which interact
with the defect mode in the crystal 
versus velocities $v_B$, $v_C$ at
fixed value $v_A = 500 {\rm m s}^{-1}$. 
The defect region is positioned unsymmetrically: 
${\bf R}_0 = (1,-3,2)$ mm. 
We chose the phase 
$\Phi = 0$ and the other parameters 
($a$ and $R_{\rm def}$) same as in
Fig.\ref{fig3}.
}
\label{fig5}
\end{figure}

\begin{figure}
\centerline{\psfig{height=7.5cm,file=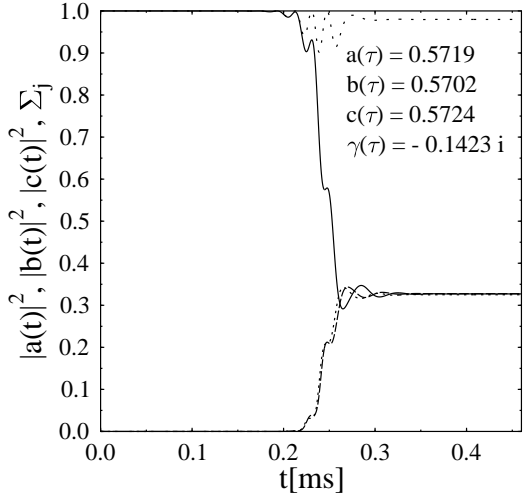}}
\begin{narrowtext}
\caption{
The time evolution of the 
atomic populations  
$|a(t)|^2$ (solid line), $|b(t)|^2$
(long dashed line),
 and $|c(t)|^2$ (short dashed line),
when the three atoms are injected
into the photonic crystal simultaneously with the velocities
$v_A = 500\ {\rm m s}^{-1}$, $v_B = 536.4\ {\rm m s}^{-1}$ 
and $v_C = 527.4\ {\rm m s}^{-1}$. All other parameters 
are the same as in Fig.\ref{fig5}.
We  see that the three atoms 
exit the crystal in a state nearly disentangled 
from the field state - we plot the sum
$|a(t)|^2 +|b(t)|^2 +|c(t)|^2$ (dotted line) which is close to unity.
}
\label{fig6}
\end{narrowtext}
\end{figure}

Let us consider the same setup as in our previous discussion except
we assume now three atoms  flying through the crystal (see
Fig.~\ref{fig1}).
These 
 three two-level Rydberg atoms ($A$, $B$ and $C$) are  injected 
into the holes at the bottom side of the crystal simultaneously. The  
atom $A$ is initially in its upper level $|e_A\rangle$ while 
atoms $B$ and $C$ are initially in their lower states $|g_B\rangle$ and
$|g_C\rangle$. The single defect mode is initially prepared in its vacuum
state $|0\rangle$.
The atoms move along the axes
of the holes and interact with the defect mode in the central region of
the crystal. The electric-field amplitude of the mode is given by
Eq.~(\ref{shape}). We consider slightly asymmetric position of the
defect mode in the crystal (the reason is explained below).
In Fig.\ref{fig5} we present plots of the final atomic populations 
versus velocities
$v_B$ and $v_C$ while $v_A$ is fixed at the value $500 {\rm m s}^{-1}$.
These plots show that adjusting atomic velocities we
can obtain required probabilities 
such that in the final state (\ref{finalstate2}) the probability  amplitude
$\gamma(\tau)$ is equal to zero, which means that the defect mode
is decoupled from the atomic system. The atoms are then in a pure
superposition state.
In particular, if we select velocities $v_B = 536.4\ {\rm m s}^{-1}$ and 
$v_C = 527.4\ {\rm m s}^{-1}$, we
obtain a final state with 
equal probabilities 
$|a(\tau)|^2 = |b(\tau)|^2 = |c(\tau)|^2 \approx 0.33$ 
(see Fig.\ref{fig6}). 
Square of $|\gamma(\tau)|$ gives the probability of the photon 
in the final state 
approximately equal to $0.02$. It means
that the atomic subsystem is in a good approximation  
decoupled from the field subsystem.

\end{multicols}
\vspace{-0.5cm}
\widetext
\begin{multicols}{2}

We have chosen an  asymmetric position  of the defect
mode with respect to the center of the crystal 
because for the symmetric position we were able to obtain the
``symmetric'' result $|a(\tau)|^2 = |b(\tau)|^2 = |c(\tau)|^2 \approx 0.33$ 
only when two of the velocities are equal. In this case we face
the problem of the collision of the atoms. 
We expect that a better choice of
the defect geometry might produce a final state more disentangled from the
field as is the case presented in Fig.\ref{fig6}.

We  see from Figs.\ref{fig5} that
variations of the final atomic populations are rather robust with respect
to changes in velocities, i.e. uncontrolled velocity fluctuations
(which in experiments can be reduced up to
$0.4\ {\rm m s}^{-1}$ \cite{Hagley}) do not deteriorate the predicted
entanglement.


\section{Conclusions}
\label{s_last}
In this paper we have shown that
atoms can be entangled  in photonic crystals  via dipole 
interaction mediated by off-resonant modes or 
via an interaction with a  single 
defect mode. 
In the first mechanism (RDDI)
the atoms can coherently exchange excitation
while only a very small part of this energy is radiated into the field.
However, this interaction might not be easy to control in an
experiment because it requires a high precision position control
of the position of the atoms.
The second mechanism (via a single resonant defect mode) is 
experimentally more promising 
because it can be realized with
currently available microwave photonic crystals and with 
highly excited Rydberg atoms.

We have shown that atoms can be prepared in pure entangled states and that 
the probability amplitudes of the 
generated superposition states of the atoms can be coherently controlled 
by varying  the velocities of the atoms 
or by varying the orientations of the atomic dipole matrix elements.

In our scheme of entanglement via defect modes in 
photonic crystals  the distance between the entangled
atoms at the exit from media depends on the size of
the media, the angle between the atomic trajectories, the
atomic velocities and the life of the atoms. For the parameters
used in this paper the distance between the entangled atoms is
of the order of tens of centimeters.

Finally, we think that investigation of dynamics of Rydberg atoms in 
photonic crystals is an interesting complement to 
current experimental cavity quantum electrodynamics. 

\end{multicols}

\newpage

\end{document}